# 4G Mobile Communication Systems: Key Technology and Evolution

Seyed Hossein Ahmadpanah

Department of Computer and Information Technology, Mahdishahr Branch, Islamic Azad University, Mahdishahr, Iran

Abdullah Jafari Chashmi

Department of Electrical and Telecommunications Engineering Technology, Mahdishahr Branch, Islamic Azad University, Mahdishahr, Iran

Majidreza Yadollahi

Department of Computer and Information Technology, Mahdishahr Branch, Islamic Azad University, Mahdishahr, Iran

*Abstract*— **With the worldwide third-generation mobile communication system gradually implemented, the future development of mobile communications has become a hot topic and evolution of the problem. This paper introduces the fourth generation mobile communication system and its performance and network structure and OFDM, software defined radio, smart antennas, IPv6 and other key technologies, and analyzes the relationship between 4G mobile communication system for mobile communications and 3G, and the evolution of communication systems do Prospect.**

## I. Introduction

Currently, the third generation mobile communication (3G) standards and norms has formed agreements, from 2001 onwards has commercially in Japan and Korea, but most countries are still operating a 2G or 2.5G mobile communication systems. The mobile communication system of the operations is mainly 2G GSM / GPRS and CDMA systems. Currently users of the data transmission rate of the mobile communication systems have become increasingly demanding, and the highest rate of actual 3G system can provide only 384kbps (although the nominal maximum rate of 2Mbps), cannot meet the actual needs of users, so the 3G system the case has not been put into large-scale commercial, domestic and foreign experts in the field of mobile communications has begun to carry out research and development work 4G (or B3G) systems. [1]

The International Telecommunication Union (ITU) as early as in September 1999 the standardization of the "third generation after" the mobile communication system under the put on the agenda included the "IMT 2000 and beyond in ITU-R work plan system ", ITU formulation is about 4G Beyond IMT-2000 (3G), and proposed to Member States to achieve 4G commercially in 2010. But now 4G is only a basic framework for it, the definition is not clear.

At present description of the fourth generation mobile communication system mainly has the following aspects:

1. the establishment of a new band (e.g. 5-8GHz or higher) on a wireless communication system, based on high-speed transmission of packet data (50Mbps or more), hosting a large number of multimedia information, it has an asymmetric uplink and downlink rate, continuous coverage area, QoS mechanisms, low bit overhead functions [2];

2. the real "global unity" (including the satellite component) communication system, network system based on the new system, or that it would be part of the new wireless network (intelligent, multi-service support, can be moved management) "wireless access" to make between various types of media, communications and network host a "seamless" connection, enabling users to freely roam seamlessly between a variety of network environments [3];

3. the "communication" system is not simply in the traditional sense, but the convergence of digital communications, digital audio / video receiver's new system (on demand) / and Internet access, the user can freely select protocols, applications and networks. Keep an application service provider (ASP) and content providers to provide independent of the operating business and content [4].

## II. Introduction For 4G mobile communication system in key technology

### A. Network structure

Currently, 4G system is still in the initial stage of the study, the relevant standard has not yet introduced, the network structure is not formed, but the network convergence trend is obvious. [5] Map of "all-IP core network" includes a whole from the IP backbone of the transport layer to the control layer, application layer. Future radio base station through the IP protocol will have the ability to directly access the "all-IP core network", the main function of the existing 2G mobile communication system switching center MSC, a home location register HLR, authentication center AUC and other network elements are by 4G or database server on the network to achieve, signaling the Internet protocol layers will gradually be replaced by the IP protocol. The entire network from the past evolution of distributed vertical tree routing structure, business differences are only reflected in the access level.

4G communication systems according to their functions can be divided into the access layer, the carrier layer and the service control layer 3 layers. Access layer allows users to use a variety of terminals through various forms of access to the 4G communication system, this part will be revolutionary evolution; carrier layer to provide QoS assurance, security management, address translation and other functions, and between the access layer interface should be open IP protocol interface; the service control layer provides the business management, loading and other functions, should also be open interface between it and the load-bearing layer, in order to provide a new third-party business applications[6].

From the foregoing description of the 4G communication system can be seen, it is a far more complex 3G communication system, its implementation needs to rely on the many emerging technologies. In the 4G communication systems used in key technology may include OFDM, software defined radio, smart antennas, and other mobile IPv6, the following description of each key 4G communication system technology.

*B. OFDM (Orthogonal Frequency Division Multiplexing)*

Due to the presence of multipath radio channel, when the data signal is transmitted on different types of radio channels, the resulting delay will cause inter-symbol interference of the received signal, particularly when the symbol rate and improve the shortened cycle time, the time delay spans more symbols, leaving such interference becomes larger. In addition, the speed increase symbol causes a corresponding increase in signal bandwidth, the bandwidth of the signal bandwidth when the associated big channel will cause frequency selective fading. Currently single-carrier modulation technique in order to minimize the use of this fading equalization, but had to increase the channel noise as the price.

Future wireless multimedia services first requires a higher data transfer rate, while also ensuring the transmission quality, which requires the use of modem technology both have a high cell rate, but also have a longer symbol period. Based on this consideration, resulting OFDM technology, it is a multi-carrier modulation technique (MCM) one. OFDM is the core technology of 4G communications network. The main idea is: The channel is divided into a number of orthogonal sub-channel, high-speed data signal is converted into parallel low-speed sub-data streams, modulation transmit on each sub channel. Quadrature signals can be used at the receiver to separate the related art, thus reducing the mutual interference (ICI) between the sub channels. Signal bandwidth of each sub channel is less than the relevant channel bandwidth, it can be regarded as flat fading on each sub-channel, which can eliminate inter-symbol interference. And because each sub channel bandwidth is only a small part of the original channel bandwidth, channel equalization becomes relatively easy [9]. OFDM technology becoming more and more attention, because OFDM has many unique advantages:

1. a high spectral efficiency, higher spectral efficiency than serial systems nearly doubled. This is important in the limited spectrum resources in wireless environments. Adjacent sub-carriers of the OFDM signal overlap with each other, in theory it can be close to the spectrum utilization NY Quist limit.

2. anti-fading ability. OFDM user information through a plurality of sub-carrier transmission signal time on each subcarrier is correspondingly long signal time compared with the rate of single-carrier system many times, so OFDM impulse noise (Impulse Noise) and the channel fast fading more resistant. At the same time, through the joint coding sub-carriers, to the frequency diversity effect between sub channels, also enhanced impulse noise and channel fast fading resistance. Therefore, if the decline is not particularly serious, there is no need to add the time domain equalizer.

3. suitable for high-speed data transmission. OFDM adaptive modulation mechanism so that the different sub-carriers may use different channels according to the situation and the background noise of different modulation methods [7]. When the channel condition is good, high efficiency modulation scheme. When channel conditions are poor, the use of anti-interference ability of modulation. Furthermore, OFDM loading algorithm uses, the system can put more focus on the good condition of the data channel is transmitted at a high rate. Therefore, OFDM technology is ideal for high-speed data transmission.

4. between the Inter-symbol interference (ISI) ability. ISI is the most important digital communication system noise interference in addition, it is additive noise is different, is a multiplicative interference. ISI causes are many, in fact, as long as the transmission channel bandwidth is limited, it will cause certain inter-symbol interference. As a result of the cyclic prefix OFDM, inter-symbol interference ability to fight strong.

In addition to the above advantages, OFDM also has three obvious shortcomings.

First, the frequency offset and phase noise sensitive. Orthogonality causes deterioration of the phase noise and frequency offset between subcarriers for each OFDM signal to noise ratio drops.

Secondly, the average power and the peak ratio (the PARR), leading to a lower transmission power efficiency of the amplifier. Because OFDM signal is composed of a plurality of independently modulated subcarrier signals are added in the synthesis, it is possible to generate a relatively large peak power, it is possible to have a greater value PARR [10]. And generally have a high value PARR transmission side high power amplifier linearity requirements raised, thereby increasing the cost of the base station and the user terminal [8].

Third, the complexity of the system modulation technology enables adaptive increased. OFDM uses adaptive modulation technique will increase the complexity of the transmitter and receiver, and when the mobile terminal moving speed reaches the vehicle, adaptive modulation techniques there is no great sense.

*C. Software Radio*

The Software-Defined radio (Software Defined Radio, abbreviated SDR), is the use of digital signal processing

technology, the programmable control common hardware platform, using software to define the achievement of some of the features of radio stations include: front-end receiver, the IF signal processing, and baseband processing [11]. That the whole radio station from the high-frequency, IF, Baseband until the control protocol part of all software programming to complete.

The basic idea of software radio is the basic hardware as its common platform to as many personal and wireless communication function is achieved through the programmable software, making it a more work-band, multi-mode, multi-signal transmission and processing radio systems. It can be said, it is a kind of software to realize wireless communication physical layer connections.

The core software radio technology is a wideband radio receiver to replace the original narrow-band receivers and wideband analog / digital and digital / analog converter as close to the antenna, so that as much of the function of the communication station can use programming software. The structure shown in Figure 1.

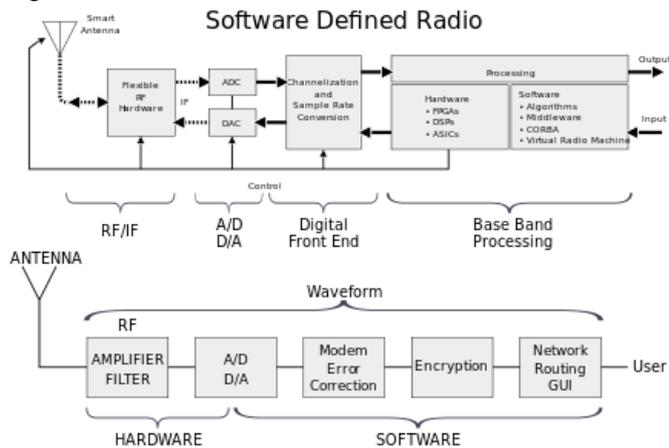

Figure 1. Software radio structure diagram

Software Radio main advantage in the following aspects:
1. General system structure, function flexible, easy to upgrade improvements. Mode by software programming changes, including a programmable radio frequency band broadband access mode and programmable signal modulation scheme[5]. It is possible to replace any channel access mode, change the modulation scheme or receive signals of different systems; business can be expanded through software tools to analyze wireless communication environment, define the desired enhanced business environment and real-time testing, upgrade and convenient.
2. provides the possibility of interoperability between different systems. Software allows mobile radio terminal for all types of air interface, you can switch between different types of traffic. A plurality of channels shares a common RF front-end and broadband A / D, D / A converter for each channel of relatively inexpensive signal processing performance.
3. Due to achieve the main function of the system through software, it is easier to adopt new signal processing methods, thereby improving the anti-interference performance.
4. a strong track new technologies have been Because it can ensure the basic hardware platform structure of the case does not change by changing the software to implement new services and new technologies, which greatly reduces the communications equipment manufacturers of new product development costs and cycles, but also reduces the operators' investments.

Realization of software radio technology needed to overcome these difficulties.
1. Design of multi-band antenna. Software radio antennas need to cover multiple bands to meet the needs of multi-channel communication simultaneously in different ways, and different radio frequencies and propagation conditions, so that the band of the antenna requires the presence of large differences, so the multi-band antenna design become one of the difficulties of software radio technology.
2. High speed A / D, D / A converter. According to NY Quist sampling theorem, the signal to recover the original signal without distortion from the sampling, the sampling frequency should be greater than 2 times the highest frequency signal. At present, the maximum sampling frequency of A / D, D / A subject to its performance limits, which can also limit the frequency modulated signal processing.
3. High-speed DSP (digital signal processor). Mainly to complete the high-speed DSP chip modem and a codec process various waveforms, it requires more computing resources and higher operation speed broadband processed by A / D, high-speed data stream D / A converted, so its chips to be further developed.

*D. Smart Antenna*

Smart antenna is defined as a multi-beam antennas or adaptive array beams does not switch. Multiband antenna using a plurality of fixed beams in a sector, and in the adaptive array antenna receiving a plurality of signals are weighted and synthesized together so that the maximum signal to noise ratio. Compared to the fixed beam antennas, in addition to the advantages of the antenna array is the high-gain antenna, but also provide a multiple diversity gain. However, they require each antenna has a receiver, but also to provide a multiple diversity gain[9].

Smart Antenna inhibit signal interference, automatic tracking and digital beam adjustment and other intelligent features, the basic working principle is adjusted according to the direction of the signal wave pattern adaptively tracking the strong signal, reduce or offset the interference signal. Smart antennas can improve the signal to noise ratio, improve the quality of the communication system, ease the contradiction between the growing shortage of spectrum resources and radio communication, reduce overall system cost, so it is bound to become a key technology 4G system. Core smart antenna is a smart algorithm, and the algorithm determines the complexity of the circuits and transient response speed, so you need to select a good algorithm intelligent control beam[10].

2G communication system currently used omnidirectional antenna is divided into portions both antennas and directional antennas, the whole cell coverage applies to 360 ° directional antenna, directional antenna used in cell division after the cell coverage. Both the shape of the area covered by the antenna is constant, and therefore the base station, the downlink to each mobile user's signal is broadcast transmission, this will inevitably lead to system disturbances and reduce the system capacity.

Smart antennas using spatial division multiple access (SDMA) technology that uses the difference in signal transmission direction, the same frequency or the same time slot, with the signal code channel distinguish dynamically change the coverage area of the signal, so that the main beam is aligned direction of the user, or the side lobe nulling aligned with the direction of interfering signals, and users can automatically track and monitor changes in the environment, provide quality uplink and downlink signal for each user, so as to suppress interference and accurately extract the effective signal purpose[8].

Therefore, smart antenna technology is more suitable for a mobile communication system having a complex wave propagation environment. In my country's 3G standard TD-SCDMA smart antenna technology is adopted.

Smart antenna has the following advantages:
1. increase system capacity. SDMA using smart antenna technology, the use of the different spatial directions channel segmentation, can use the same frequency without interference in different channels at the same time, thereby increasing system capacity.
2. reduce system interference. Smart antenna technology to beam side lobes or interference signal nulling aligned direction, it is possible to effectively suppress interference.
3. expand the coverage area. Because smart antenna with adaptive beam steering function, as compared with the conventional antenna, transmission power in the same conditions, using smart antenna signal can be transmitted to greater distances, thereby increasing coverage.
4. reduce the cost of system construction. Because smart antenna technology to expand the coverage area, the number of base station construction can be reduced, reducing the operator's construction costs. The main disadvantage of smart antenna technology is the use of it will increase the complexity of communication systems and components offer higher performance requirements.

*E. IPv6*

4G communication system is selected by using a transmission data stream based on all the way IP packets, so the IPv6 protocol will become the core technology of next generation networks[6]. Select the IPv6 protocol is mainly based on two considerations, the point is enough address space, another point is to support mobility management, both of which are not available in IPv4. In addition, IPv6 can also provide better QoS than the IPv4 and ensure better security. Since the bearer network is an IP network, future mobile terminals necessarily need to have a unique IP address as the identity. Currently used IPv4 address length of only 32bit, its IP address resources are expected to be depleted around 2006. And having up to 128bit IPv6 address space, that is up to 2128 addresses, it is possible to solve the problem of insufficient resources to address.

4G communication system is selected by using a transmission data stream based on all the way IP packets, so the IPv6 protocol will become the core technology of next generation networks. Select the IPv6 protocol is based on the following considerations:
1. a huge address space. Within a foreseeable period of time, it can provide a globally unique address for all network devices can imagine.
2. automatic control. IPv6 there is another basic feature is that it supports two kinds of stateless and stateful address auto-configuration mode. Stateless address auto-configuration mode is the key to address. In this way, you need to configure the node address of the use of a neighbor discovery mechanism to obtain a local connection address. Once you get this address, it uses another mechanism to plug and play, without any manual intervention, to obtain a globally unique routing address. Stateful configuration mechanisms such as DHCP (Dynamic Host Configuration Protocol), requires an additional server, and therefore need a lot of additional operations and maintenance.
3. quality of service. Quality of Service (QoS) comprises several aspects. From a protocol perspective, IPv6 and the current IPv4 provides the same QoS, but the advantages of IPv6 is reflected in providing different services. These advantages from IPv6 header newly added field "stream flag." With the 20-bit field, during transmission, each node can be identified in Iran and a separate deal with any IP address stream. Although the exact application of this stream flag has not yet worked out the relevant standards, but in the future the new billing system which is used based on service levels.
4. mobility. Mobile IPv6 (MIPv6) to provide greater flexibility in terms of new features and services. Each mobile device has a fixed home address (home address), irrespective of the position of the address and the device is currently connected to the Internet[3]. When the device is used outside of the home place to provide mobile node's current location information through a care-of address (care-of address). Mobile device every time you change location, it will be forwarded to the address to the home address and its corresponding communication node. Place outside the home, the mobile device transmits data packets are usually transmitted in the address in the IPv6 header as the source address.

III. RELATIONSHIP OF 4G MOBILE COMMUNICATION SYSTEM WITH 3G SYSTEM

From the foregoing description 4G communication system can be seen, it can be better than the 3G system, providing more convenient multimedia communication services based in the future 4G system will replace the 3G system is an inevitable trend in the development of communication systems. However, the development of 3G systems is also essential.

First, the construction of 3G systems is actually able to use a wider range of multimedia services play a role in the future market development, as the first generation analog systems to foster mobile business users, GPRS systems for data 3G services to cultivate the same. Users of the new service from the initial recognized accepted until the last widespread use is a lengthier process that requires operators from a simple business to diversify the business gradually provided by first attract high-end users to gradually spread to the middle and low end user, allowing users to selectively use from occasional to extensive natural application, it all depends on market development[11].

Secondly, each new technology from the initial concept to the proposed technical difficulties breakthrough to build a test network, and then to the final product of the market is a lengthy application process. In this process, the user traffic types and content requirements are gradually improving communication service market could not have been in a wait state, waiting for a new powerful system to solve all problems at once, but rather in the development of the market each stage must have a communication system can be adapted with support. Thus, at the beginning of the development of data and multimedia services, to meet the user to build a simple 3G communication system requirements is a must. For equipment developer, the construction of 3G systems to help them find the problem from the actual operation of 3G networks in order to explore a better solution, so as to provide valuable guidance for the future construction of 4G systems.

Third, from the smooth evolution of the network, who, 3G system is also an essential stage. The current 2G systems from the access network to the core network are all type of circuit, and future 4G systems are integrated from the access to the core network all-IP architecture. Evolved from a complete circuit of the system domain system-wide configuration of a packet-based IP is a step-jump, either from the operator's point of view or from the user's point of view, this change is significant.

For operators, network construction in addition to the non-smooth evolution will bring great investment risk than for investments already existing network will also be in vain. For users, it will be necessary to replace the face of the mobile terminal helpless choice, in this case will result in a large number of users to reselect carriers and network operators will also bring immeasurable loss[2]. From the target 4G is proposed to solve the 3G system deficiencies, provide a sound technical system idealized, although 3G to 4G for both the network or terminal and cannot be completely smooth evolution, but in the middle of the 3G system can play a connecting link pivotal role, to become all-IP network evolution indispensable part of Figure 2. from 2G to 4G evolution shown.

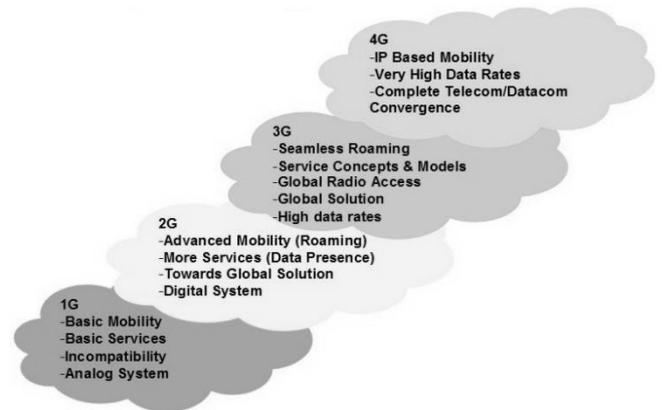

Figure 2. Evolution from 2G to 4G

From these aspects, although 3G roaming systems in communication speed, heterogeneous systems there are all kinds of undesirable places, but it is not a bubble, but inevitable stage of development of communication networks. And the construction of 4G systems will also be built on the 3G users already cannot meet the future demand for multimedia services on the basis of carried out[6].

According to the development from 1G to 3G, we can see the development cycle of the communication system is usually 10 years, so it is generally believed from 20lO began the era of 4G systems will come. At present, Iran's 3G system construction has not yet started, but the 3G licenses may be issued at the end of this year or early next year, once the 3G licenses may soon have an orgy 3G network construction. Taking into account the actual situation of Iran's communication development is still the main demand voice services, 3G systems there are some limitations on multimedia services, the practical effect is to 4G 3G multimedia services market development, the future 4G revolutionary system equipment Evolution of the potential impact on investment and other factors, it is considered the mainstream of 3G services should still be the main voice and data services, multimedia age still waiting for the construction of 4G system, so our major carriers in the choice of the timing of the construction of 3G systems and care should be taken decision on the scale of construction.

IV. CONCLUSION

Nearly 25 years of development can be seen in mobile communications, after a technical standards produced (even before its commercial), technical defects or limitations already show it; and when this technology on the market to peak, it stimulates market demand is out of the business beyond its ability to supply a new generation of technology came into being. The first generation is so, the second generation, too, the third generation, fourth generation systems will necessarily so. Moreover, With the rapid development of electronic information technology, the generation of technology market increasingly short life expectancy, which is the logic of historical development. Therefore, the standardization of the current third-generation system is nearing completion, the application system is about to launch, on next generation (fourth generation) mobile communication system is imperative. Now 4G dawn has emerged, we believe that 4G communications world, people's lives will be more exciting.